# Observation of oscillatory magnetoresistance periodic in *1/B* and *B* in $Ca_3Ru_2O_7$

V. Durairaj[1], X. N. Lin[1], Z.X. Zhou[2], S. Chikara[1], E. Ehami[1], A. Douglass[1], P. Schlottmann[3] and

G. Cao[1*]

[1]Department of Physics and Astronomy, University of Kentucky, Lexington, KY 40506

[2] Oak Ridge National Laboratory, Oak Ridge, TN 37831

[3] National High Magnetic Field Laboratory, Tallahassee, FL 32310

We report magnetoresistance oscillations in high magnetic fields, B, up to 45 T and over a wide range of temperature in the Mott-like system $Ca_3Ru_2O_7$. For B rotating within the *ac*-plane, slow and strong Shubnikov-de Haas (SdH) oscillations periodic in 1/B are observed for T≤1.5 K in the presence of metamagnetism. These oscillations are highly angular dependent and intimately correlated with the spin-polarization of the ferromagnetic state. For B||[110], oscillations are also observed *but periodic in B* (rather than 1/B) *which persist up to 15 K*. While the SdH oscillations are a manifestation of the presence of small Fermi surface (FS) pockets in the Mott-like system, the B-periodic oscillations, an exotic quantum phenomenon, may be a result of anomalous coupling of the magnetic field to the $t_{2g}$-orbitals that makes the extremal cross-section of the FS field-dependent.

Quantum oscillations, a manifestation of oscillations of the density of states due to Landau levels, provide a direct probe to Fermi surface. This probe is particularly important for investigations of correlated electron systems whose quantum oscillations are experimentally observable. $Ca_3Ru_2O_7$ belongs to a class of new highly correlated electron materials which is rich in novel phenomena. The central feature of 4d-electron based materials is the extension of their orbitals that leads to competing energies including crystalline fields, Hund's rule interactions, spin-orbit coupling, p-d hybridization and electron-lattice coupling. It is the interplay of the electron spin with the orbital and lattice degrees of freedom that leads to exotic new properties, which are particularly susceptible to perturbations such as an external magnetic field and slight structural changes. The tilting of $RuO_6$ octahedra determines the overlap between the three $t_{2g}$ orbitals ($d_{xy}$, $d_{zx}$ and $d_{yz}$) of neighboring corner-shared octahedra, and small changes in this tilting can result in drastic qualitative changes giving rise to complex phenomena [1, 2]. The long range spin and orbital order (OO) of $Ca_3Ru_2O_7$ strongly depends on the orientation of B, hence, can be effectively manipulated and probed with a magnetic field. In this Letter, we present inter-plane or c-axis magnetoresistivity, $\rho_c$, with B (i) rotating within the *ac*-plane and (ii) aligned with the [110] direction (diagonal direction in the *ab*-plane) for temperatures ranging from 0.4 K to 15 K and magnetic fields B up to 45 T. The results reveal slow yet strong SdH oscillations (periodic in 1/B) in the *ac*-plane with frequencies ranging from 30 to 117 T in the presence of metamagnetism, which leads to a ferromagnetic (FM) state. These frequencies are highly angular dependent and intimately correlated with the FM state. For B||[110], *unusual* oscillations in the magnetoresistance are observed with *periodicity in B* (rather than 1/B) persisting up to 15 K with the amplitude of the oscillations slowly decreasing with increasing temperature. While observations of the SdH effect are uncommon in oxides and unexpected in Mott systems such as $Ca_3Ru_2O_7$ because of large resistivity and the carrier mean-free path comparable with the lattice spacing, the oscillatory behavior periodic in B, to the best of our knowledge, has not been seen in bulk materials before, reflecting unusual coupling of the Fermi surface to magnetic field. The aim of this Letter is to present this exotic new physics.

The crystal structure of $Ca_3Ru_2O_7$ is severely distorted by a tilt of the $RuO_6$ octahedra, which projects primarily onto the *ac*-plane (153.22°), and only slightly affects the *bc*-plane (172.0°) [3]. These crucial bond angles directly impact the band structure and are the origin of the anisotropic properties. For low fields, $Ca_3Ru_2O_7$ undergoes an antiferromagnetic (AFM) transition at $T_N$=56 K while remaining metallic, and then a Mott-like transition at $T_{MI}$=48 K [3-14]. This transition is accompanied by an abrupt shortening of the *c*-axis lattice parameter below $T_{MI}$ [6]. Such magnetoelastic coupling results in Jahn-Teller distortions of the $RuO_6$ octahedra, lowering the $d_{xy}$ orbitals relative to the $d_{zx}$ and $d_{yx}$ orbitals with an orbital distribution of ($n_{xy}$, $n_{zx}/n_{yz}$)=(2,2). Consequently, a phase with AFM *and* OO can occur, explaining the nonmetallic behavior for T<$T_{MI}$ and B<$B_c$ (metamagnetic transition). This scenario is consistent with Raman-scattering studies of $Ca_3Ru_2O_7$, which probe magnon, phonon and thus orbitals [11, 12, 15].

Shown in Fig.1a is the B-dependence of the resistivity for the current along the *c*-axis, $\rho_c$, (right scale) for T=0.4 K and 0≤B≤45 T with B||*a*-, *b*- and *c*-axis. A central feature of this system is that the resistivity is extraordinarily sensitive to the orientation of B. *For B||a-axis (magnetic easy-axis)* $\rho_c$ shows an abrupt drop by an order of magnitude at 6 T corresponding to the first-order metamagnetic transition leading to the FM state with a saturated moment, $M_s$, of 1.8 $\mu_B$/Ru or more than 85% polarized spins (see left scale in Fig.1a). The reduction of $\rho_c$ is attributed to the coherent motion of electrons between Ru-O planes separated by insulating Ca-O planes, an effect similar to spin-filters. $\rho_c$ increases linearly with increasing B by more than 30% from 6 to 45 T. Since $B^2$-dependence is expected for regular metals (quantization of orbits), a linear variation of $\rho_c$ over such a wide range of B is likely to arise from orbital $t_{2g}$ degrees of freedom, which couple to B directly and may modify the dynamics of the electrons. *For B||b-axis (magnetic hard-axis)* there is no spin-flop transition and the system remains AFM. However, for B||*b*-axis, $\rho_c$ rapidly decreases by as much as three orders of magnitude at $B_c$=15 T, i.e., two orders of magnitude more than the drop for B||*a*, where the spins are nearly fully polarized [16]. Hence, in $Ca_3Ru_2O_7$ a colossal magnetoresistance is achieved only when a FM state with full spin polarization is *avoided*. The FM state stabilizes OO, suppressing the hopping of the electrons. This in turn makes the FM state the least favorable for conduction [16]. This behavior is fundamentally different from that of all other colossal magnetoresistive materials that are primarily driven by spin polarization [17]. *For B||c-axis,* $\rho_c$ displays slow SdH oscillations in the absence of the metamagentism, signaling the existence of very small Fermi surface cross sections. Local density approximation calculations [18] for $Sr_3Ru_2O_7$, which shares common aspects with $Ca_3Ru_2O_7$, find the Fermi surface very sensitive to small structural changes. In particular, the $d_{xy}$ orbitals give rise to small lens shaped Fermi surface pockets. The observed oscillations must then be associated with the motion of the electrons in the *ab*-plane, i.e. with the $d_{xy}$ orbitals. The oscillations in $\rho_c$ correspond to extremely low frequencies, $f_1$ = 28 T and $f_2$=10 T, which, based on crystallographic data [3] and the Onsager relation $F_0=A(h/4\pi^2 e)$ (*e* is the electron charge, *h* Planck constant), correspond to a cross-sectional area of only 0.2% of the first Brillouin zone. From the *T*-dependence of the amplitude, the cyclotron effective mass is estimated to be $\mu_c$=(0.85 ± 0.05) $m_e$. In addition, the Dingle temperature, $T_D=h/4\pi^2 k_B \tau$, a measure of scattering, is estimated to be 3 K, comparable to those of good organic metals.

The field dependence of $\rho_c$ (on a logarithmic scale) for B rotating in the *ac*-plane (B||*a*-axis, $\theta=0°$ and B||*c*-axis, $\theta=90°$) at T=0.4 K is shown in Fig. 1b for B ranging from 11 T to 45 T. Similar data for T=0.6 and 1.5 K are not displayed here. The metamagnetic transition $B_c$ occurs at 6 T for B||*a*-axis, and increases with increasing θ, i.e. as B rotates towards to the *c*-axis. The striking finding is that strong SdH oscillations are qualitatively different for $11°≤θ<56°$ and $56°≤θ≤90°$. It is then likely that *θ=56° marks the onset of the melting of the OO* state as B rotates further away from the easy-axis of magnetization (*a*-axis). This destabilizes the FM state, and thus the OO state via direct coupling to the field or the spin-orbit interaction. (This is only possible perturbatively, because the spin-orbit interaction is quenched by crystalline fields). Consequently, the electron mobility increases drastically, explaining the largely enhanced conductivity for $56°≤θ≤90°$.

For $θ < 56°$ the strong oscillations occur only for $B>B_c$ and with frequencies significantly larger than the ones previously observed for B||*c*-axis. For clarity, Fig. 1c exhibits $\rho_c$ on a linear and enlarged scale for $B>B_c$. For $0°<θ<56°$ and $B>B_c$, $\rho_c$ increases with both B and θ, and displays oscillatory behavior only for $11° ≤ θ < 56°$. While the extremal orbits responsible for the oscillations are facilitated by the FM state, it is remarkable that no oscillations are seen when θ=0 (B||*a*-axis) where the FM state is fully established at $B_c$=6 T. In contrast, no oscillations were discerned for B rotating within the *bc*-plane at B up to 45 T [19]. The bumps seen in ρ for B||*b* (Fig.1a) are not oscillatory at higher B. The *bc*-plane is perpendicular to the easy axis of magnetization and has no FM component [5, 6, 19], suggesting a critical link of the SdH oscillations to the fully polarized FM state. The FM and the different projections of the tilt angles of the $RuO_6$ octahedra onto the *ac*- and *bc*-planes [3] are expected to affect the Fermi surface.

On the other hand, for $56^o \leq \theta \leq 90^o$, the oscillations disappear for $B>B_c$ but are present for $B<B_c$, accompanying the much more conducting phase at high fields, as shown in Figs. 1b and 1c. The frequency of the oscillations seen for $B<B_c$ remains essentially unchanged with $\theta$ for $65^o \leq \theta \leq 90^o$. Since the $d_{xy}$ orbitals are believed to be responsible for the oscillations B||c-axis, the nearly constant frequency upon tilting of B suggests that the oscillations in the absence of the metamagnetism originate from a nearly spherical pocket of the same $d_{xy}$ orbitals. Conversely, the oscillations for $11^o \leq \theta < 56^o$ and $B>B_c$ could be associated with a configuration of the FM state and ordered $d_{zx}$ and/or $d_{yz}$ orbitals. These orbitals offer only limited electron hopping (as confirmed by a larger resistivity), and thus lower density of charge carriers and longer mean free path which in turn facilitates electrons to execute circular orbits.

Fig. 2 shows the amplitude of the SdH oscillations as a function of 1/B for several representative $\theta$ at (a) T=0.4 K and (b) 1.5 K. The SdH signal is defined as $\Delta\rho/\rho_{bg}$, where $\Delta\rho=(\rho_c - \rho_{bg})$ and $\rho_{bg}$ is the background resistivity. $\rho_{bg}$ is obtained by fitting the actual $\rho_c$ to a polynomial. The oscillations are strong and slow, and their phase and frequency shift systematically with changing $\theta$. The oscillations vanish for $\theta \geq 56^o$, suggesting that the extremal cross-section responsible for the oscillations is highly susceptible to the orientation of B. SdH oscillations are usually rather weak in metals [20]; the remarkably strong oscillatory behavior for $11^o \leq \theta < 56^o$ may arise from an extremal orbit with a flat dispersion perpendicular to the cross-section, so that a large constructive interference can occur. With further increasing $\theta$ ($\geq 56^o$), the impact of B on the Fermi surface becomes even more dramatic and the closed orbit is no longer observed. The closed orbit is possibly replaced by open ones that do not contribute to oscillations.

Fig.3 illustrates the angular dependence of the SdH frequency for T=0.4 K and 1.5 K (left scale) and the metamagnetic transition $B_c$ (right scale). The unusual feature is that the frequency is temperature-dependent, increasing about 15% when T is raised from 0.4 K to 1.5 K. The frequency for $B>B_c$ rapidly decreases with increasing $\theta$ and reaches about 45 T in the vicinity of $\theta=56°$, whereas the frequency for $B<B_c$ stays essentially constant for $\theta>56°$. The oscillations become difficult to measure in the vicinity of $B_c$. This is expected if $B_c$ is associated with the melting of OO. The frequencies for $B>B_c$ are significantly larger than those for $B<B_c$, suggesting the former oscillations either originate from different electron orbits or a restructured Fermi surface. The angular dependence of $B_c$, on the other hand, is rather weak for $\theta<56°$ but becomes much stronger for $\theta>56°$. Note that $\rho_c$ displays a weak plateau at high magnetic fields and $\theta$ approaching 90°, which disappears for B||c-axis because the FM state is no longer energetically favorable. Such an inverse correlation between the frequency and $B_c$ reinforces the point that the FM state reconstructs the Fermi surface and facilitates the oscillatory effect. It also suggests that $\theta=56°$, which marks the simultaneous crossover in both the frequency and $B_c$, defines the borderline between the FM-OO state ($\theta<56°$) and the orbital degenerate (OD) state ($\theta>56°$).

For B||[110], $\rho_c$ also shows oscillations in the magnetoresistance as displayed in Fig.4. The striking behavior is that these oscillations are periodic in B, instead of 1/B, with a period of $\Delta B=11$ T and persistent up 15 K. This highly unusual observation is corroborated by plotting the data both as a function of 1/B and B (see panels 4a and 4c). The oscillations die off rapidly if B slightly departs from the [110] direction (within ±5°).

Oscillations in the magnetoresistivity periodic in 1/B (SdH effect) are a manifestation of the constructive interference of quantized extremal orbits of Fermi surface cross-sections perpendicular to the field. Due to the Pauli principle the electrons are bound to follow the Fermi surface. The projection of the real space trajectory of a free electron onto a plane perpendicular to **B** reproduces the k-space trajectory rotated by $\pi/2$ and scaled by a factor $c\hbar/|e|B$. Hence, trajectories with constructive interference in real space are expected to be periodic in B rather than 1/B (the frequency is proportional to the cross-sectional area in reciprocal space, so that the relation to the real space is $B^2$). Oscillations in the magnetoresistivity periodic in B are realized in some mesoscopic systems and always related to finite size effects. Examples are (i) the Aharanov-Bohm (AB) effect [20-22], (ii) the Sondheimer effect [20, 23], and (iii) the edge states in quantum dot [24]. Each of the cases involves a geometrical confinement. The AB interference occurs when a magnetic flux threading a metallic loop changes the phase of the electrons generating oscillations in the magnetoresistance and is observed only in mesoscopic conductors [20-22], but not in bulk materials. The Sondheimer effect requires a thin metallic film with the wavefunction vanishing at the two surfaces. The thickness of the film has to be comparable with the mean free path [20, 23]. This gives rise to boundary scattering of the carriers that alters the free electron trajectories and the possibility of interference. Finally, the edge states require a quantum Hall environment with real space confinement [24].

Since the bulk material has no real space confinement for the orbits of the carriers, the most likely explanation for the periodicity as a function of B is a Fermi surface cross-section that changes with field. The $t_{2g}$-orbitals have off-diagonal matrix elements with the orbital Zeeman effect, and hence couple directly to the magnetic field. Consequently, `the magnetic field could lead to a dramatic change of the Fermi surface if it points into a certain direction. Note that the pockets involved are very small (low frequencies as a function of 1/B) and susceptible to external influences. If there is more than one conducting portion of the Fermi surface, occupied states can be transferred from one pocket to another with relatively small changes in the external parameters. This is also consistent with the 15% of change in the frequency when T is raised from 0.4 K to 1.5 K shown in Fig. 3. Indeed, the amplitude of the oscillations follows the Lifshitz-Kosevich behavior expected for SdH oscillations (see Fig.4e). It is noted that the AB effect at finite T would show the same amplitude dependence [25].

In summary, the observations of the magnetoresistance oscillations in $Ca_3Ru_2O_7$ periodic both in B and 1/B reflect the crucial dependence of the quantized orbits on the orientation of B. The novel phenomena highlight the critical role of the orbital degrees of freedom embodied via the coupling of the $t_{2g}$-orbitals to the magnetic field.

**Acknowledgements** The work was supported by NSF grant No.DMR-0240813. P.S. acknowledges the support by NSF grant No. DMR01-05431 and DOE grant No. DE-FG02-98ER45707.

*Corresponding author:cao@uky.edu

14. Single crystals were grown using both flux and floating zone techniques and characterized by single crystal x-ray diffraction, SEM and TEM. All results indicate that the single crystals are of high quality. The highly anisotropic magnetic properties of $Ca_3Ru_2O_7$ are used to determine the magnetic easy *a*-axis and to identify twinned crystals that often show a small kink at 48 K in the *b*-axis susceptibility. There is *no difference* in the magnetic and transport properties and Raman spectra of crystals grown using flux and floating zone methods. Our studies on oxygen-rich $Ca_3Ru_2O_{7+\delta}$ show that the resistivity for the basal plane displays a downturn below 30 K, indicating brief metallic behavior. The metallic behavior can be readily induced by other impurity doping, such as La [3].

**Captions:**

**Fig.1.** (a) The *c*-axis resistivity $\rho_c$ at T=0.4 K (right scale) and magnetization M (left scale) at 2 K as a function of B ranging from 0 to 45 T for B||*a*-, *b*- and *c*-axis; (b) $\rho_c$ for B rotating in the *ac*-plane with θ=0 and 90° corresponding to B||*a* and B||*c*, respectively; (c) Enlarged $\rho_c$ on a linear scale for clarity. Note that the range of B is from 11 to 45 T in (b) and (c).

**Fig.2.** The amplitude of the SdH oscillations defined as $\Delta\rho/\rho_{bg}$ as a function of inverse field $B^{-1}$ for various θ and for T=0.4 and 1.5 K.

**Fig.3.** The angular dependence of the frequency (solid circles for B>$B_c$, and empty circle for B<$B_c$) for T =0.4 K and 1.5 K (left scale) and the metamagnetic transition $B_c$ (solid squares) (right scale). Note that the dash line at 56° marks the melting of OO as θ increases.

**Fig.4.** (a) The amplitude of the quantum oscillations as a function of B for B||[110] and T=0.5 K and (b) for various temperatures up to 15 K. (c) The amplitude of the quantum oscillations as a function of inverse field $B^{-1}$ for B||[110] and T=0.5 K and (d) for various temperatures up to 15 K. (e) The amplitude of the quantum oscillations (QO) as a function of temperature.

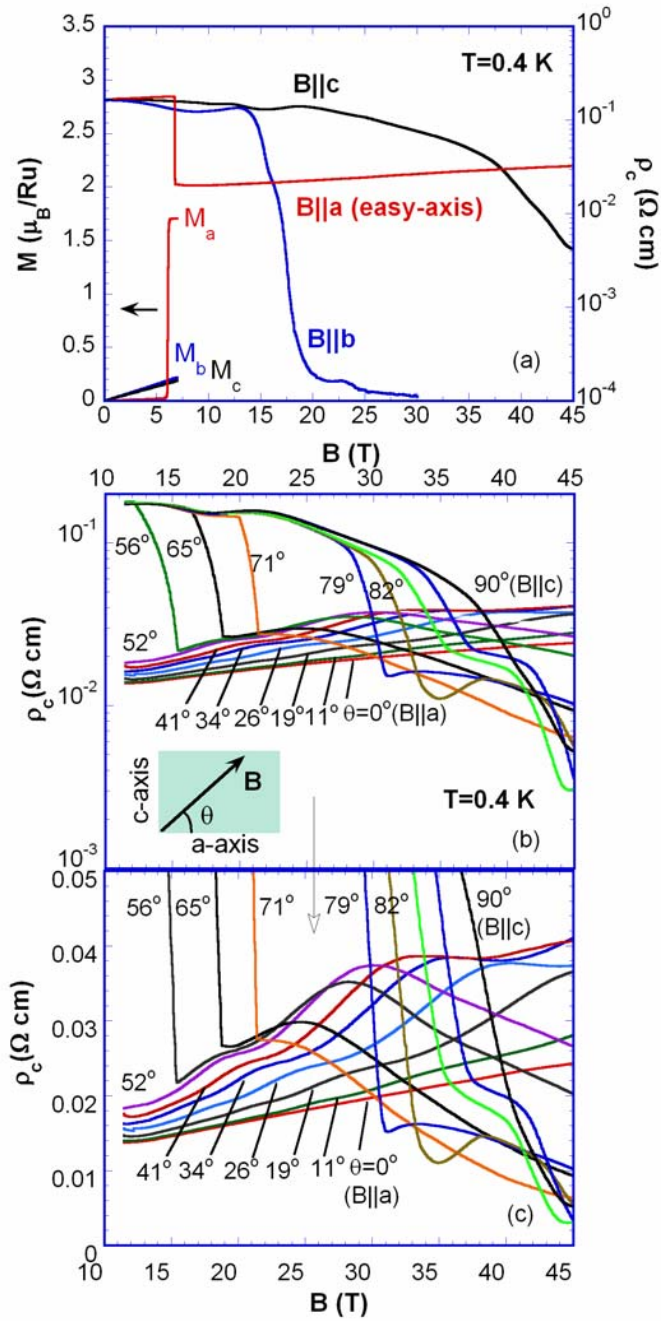

Fig. 1

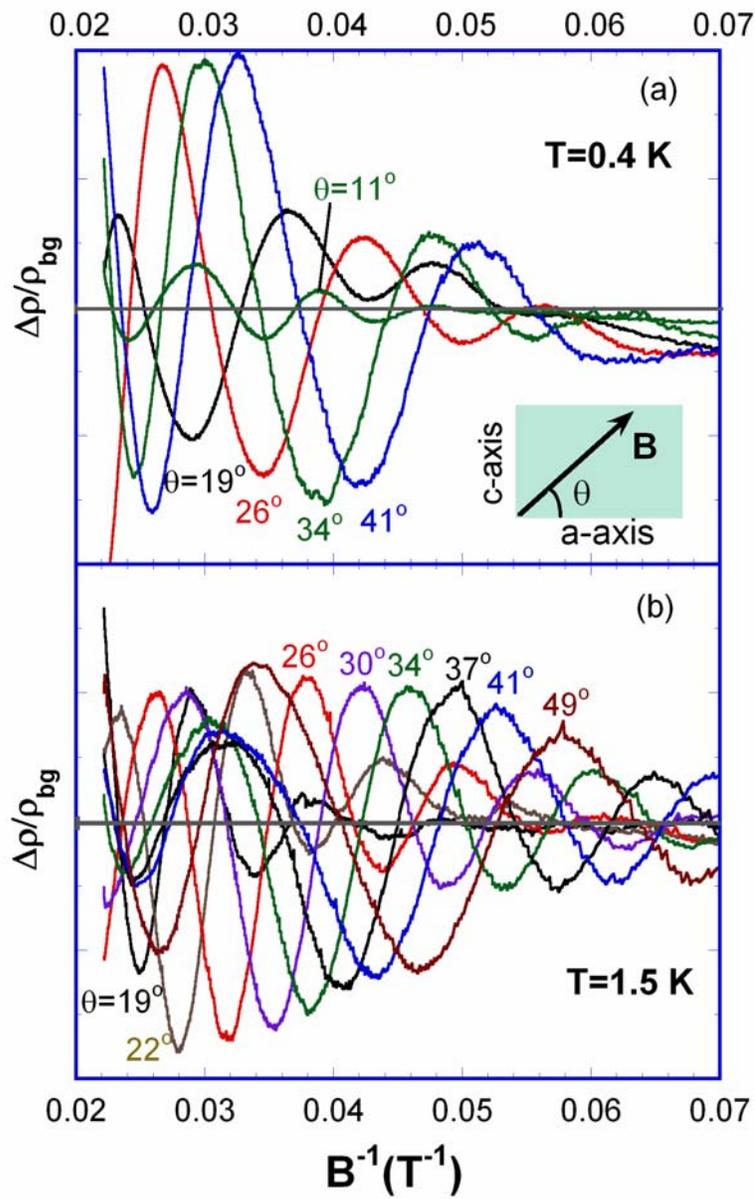

Fig.2

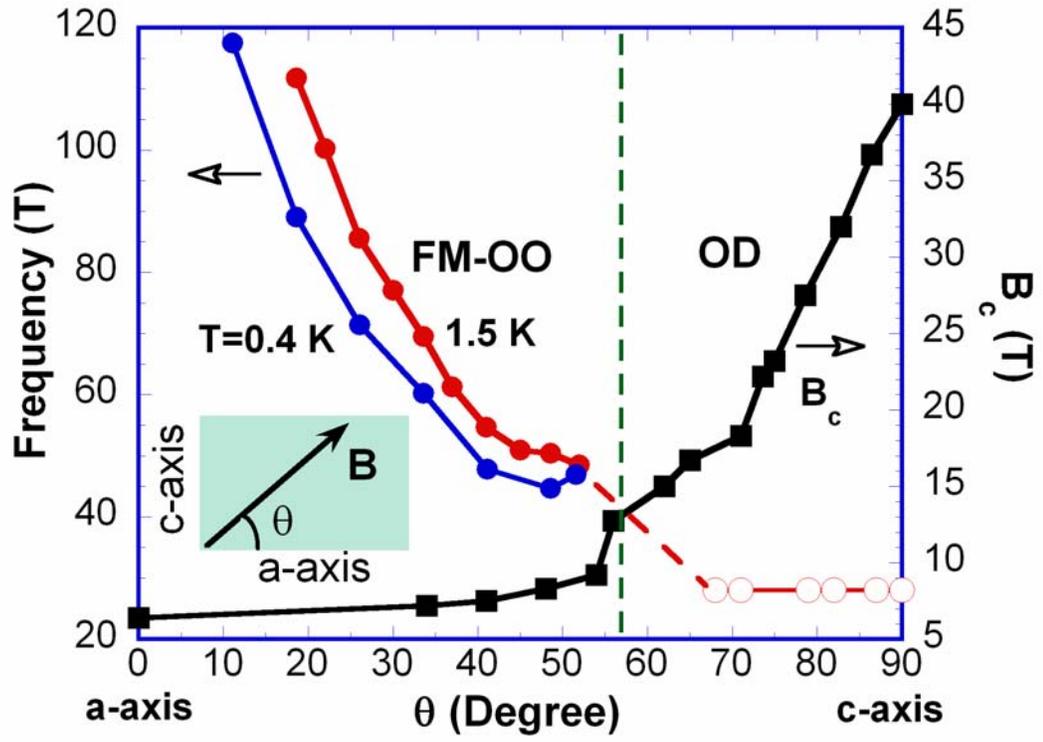

Fig.3

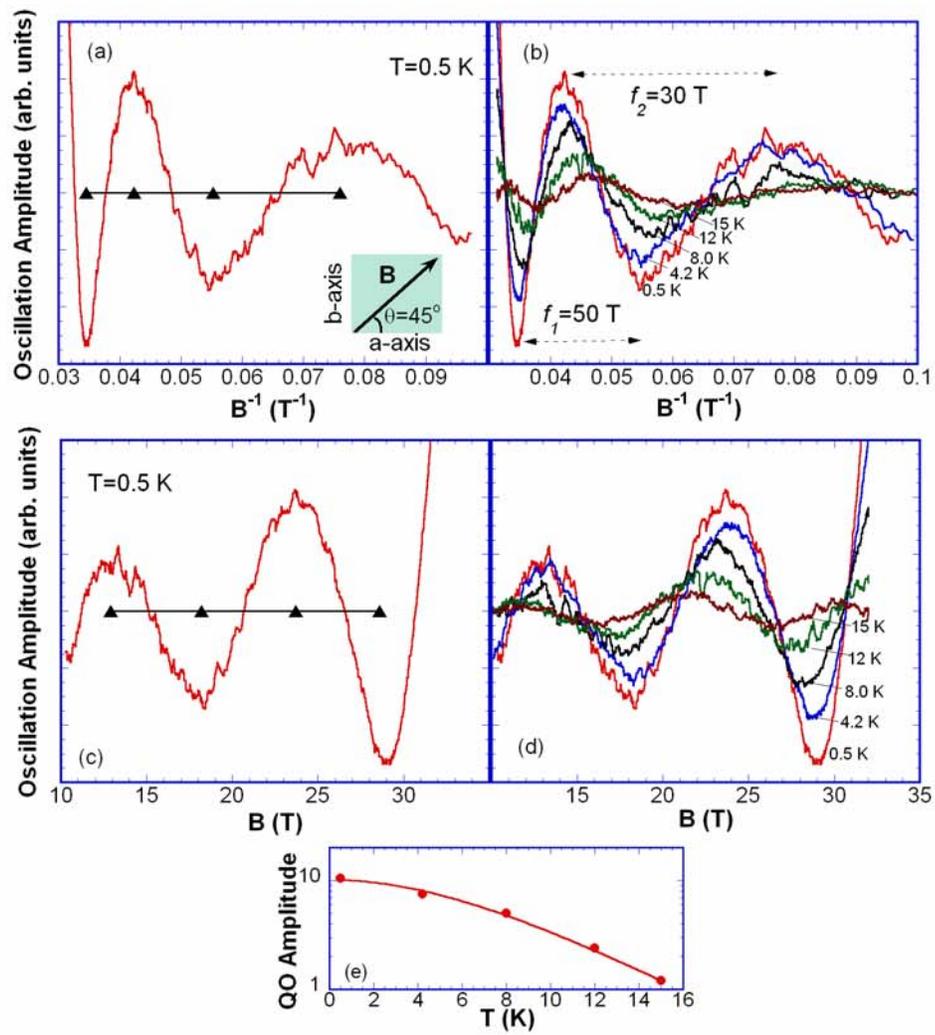

Fig.4